\documentclass[aps,superscriptaddress,prl,twocolumn,floatfix,showpacs]{revtex4}

\usepackage{graphicx}
\usepackage{amsmath, amssymb}
\graphicspath{{Figs/}}

\begin{document}
\title{Gas-Mediated Impact Dynamics in Fine-Grained Granular Materials } 
\author{John R. 
Royer}
\affiliation{James Franck Institute and Department of Physics, 
The University of Chicago, Chicago, IL 60637}
\author{Eric I. 
Corwin}
\affiliation{James Franck Institute and Department of 
Physics, The University of Chicago, Chicago, IL 60637}
\author{Peter 
J. Eng}
\affiliation{James Franck Institute and Department of 
Physics, The University of Chicago, Chicago, IL 
60637}
\affiliation{Consortium for Advanced Radiation Sources, The 
University of Chicago, Chicago, IL 60637}

\author{Heinrich M. 
Jaeger}
\affiliation{James Franck Institute and Department of 
Physics, The University of Chicago, Chicago, IL 60637}
\date{ \today}

\begin{abstract}
Non-cohesive granular media exhibit  complex responses to sudden impact that 
often differ from those of ordinary solids and liquids.  We 
investigate how this response is mediated by the 
presence of interstitial gas between the grains.  Using 
high-speed x-ray radiography we track the motion of a steel sphere 
through the interior of a bed of fine, 
loose granular material.  We find a crossover from nearly 
incompressible, fluid-like behavior at atmospheric pressure to 
a highly compressible, dissipative response once most of the gas is 
evacuated.  We discuss these results in light of recent proposals for 
the drag force in granular media.   
\end{abstract}

\maketitle

Studies of the impact of 
solid objects into granular beds have a long history, with first 
systematic work dating back to the 18th century \cite{Ro42,  Po39}. The topic reemerged 
in the 1960's with attempts to characterize the strength 
of the lunar surface \cite{RRS63, GTBZM99} and, recently, with 
investigations of cratering \cite{WHHB03, BW04, UAOD03,  Cea04, 
LRBM04,  HPLLC05, AKD05, TV05, KD07} and granular jet formation \cite{TS01, Lea04, Rea05, CBMPL07}. 
One unresolved issue remains the form for the drag force 
experienced by the impacting solid as it moves through the 
granular medium.  Various competing force laws have been proposed, 
based on scaling laws relating the penetration depth to the 
impact energy or momentum \cite{UAOD03, BW04}, or  on direct 
measurement of the trajectory of the impacting object \cite{LRBM04, Cea04,
HPLLC05, AKD05, KD07}.  However, none of these take into account the interaction between the solid grains inside a granular bed and the 
surrounding gas.

This interaction 
 is known to play a significant role in many 
situations where the bed is externally forced, especially 
for small grain sizes, when the bed's gas permeability becomes 
sufficiently low to sustain a pressure gradient that can compete with the weight of the material. The resulting
feedback between grain motion and ambient gas flow gives rise to complex dynamics not only when a bed is fluidized by direct gas 
injection \cite{Ja00}, but also in many vibrated granular systems  \cite{Gu75, PVB95, 
MCKNJ04, BKS02}.  There have been indications that the morphology of craters differs for small grain sizes due to interstitial gas flows \cite{WHHB03}.  Also,  for fine powders, the size and shape of granular jets ejected upward after impact was found to depend on the presence of interstitial gas \cite{Rea05}.  However, with one exception \cite{CBMPL07}, the role of gas-grain interactions in determining the trajectory of impacting solids has not been investigated in detail. Here we 
demonstrate that ambient gas inside a granular bed 
strongly affects the impact dynamics  and show 
how this alters the drag force in a fine-grained 
granular medium.  

Previous experiments
used either two-dimensional (2D) set-ups \cite{Cea04} or indirect methods to gauge the motion of a projectile inside a 3D bed  \cite{BW04, UAOD03, 
LRBM04,  HPLLC05, AKD05, KD07}.  In contrast, our approach is based on 
high-speed x-ray imaging which gives direct, time-resolved access to 
the dynamics in the bed interior and also allows us to extract 
local changes in the bed packing density. 

\begin{figure}
 
\includegraphics[width=7cm]{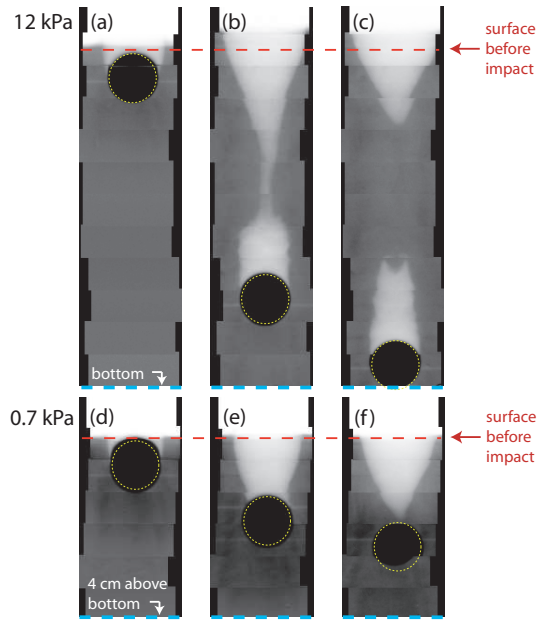}
\caption{(Color Online).  Pressure dependence of 
impact dynamics.  Composite x-ray images at 12 kPa  (a) 5 ms, 
(b) 40 ms and (c) 57 ms after impact.  Images at 0.7 kPa 
 (d) 5 ms, (e) 12 ms and (f) 28 ms after impact. }
\label{fig:xray}
\end{figure}   

For the experiments reported 
here, a steel sphere (diameter $D_s = $ 12mm) was dropped from a height of 0.34 m into an 85 
mm deep bed of boron carbide 
($\textrm{B}_4\textrm{C}$) particles (50 $\mu$m avg. diameter).  $\textrm{B}_4\textrm{C}$, which is non-spherical,  was chosen to optimize the x-ray transmission; separate experiments, studying jet formation in a variety of different media, indicate that grain shape is not a critical parameter \cite{Rea05, Rea07}.   The bed was contained in a 
cylindrical tube with 35 mm inner diameter.  Before 
each drop the bed was aerated by dry nitrogen entering 
through a diffuser built into the bottom of the container.   By 
slowly turning off the nitrogen flow, the packing  fraction $\phi = V_g/V_{tot}$, where $V_g$ is volume occupied by grains and $V_{tot}$ is the total volume of the bed,
 was adjusted  before each drop to a value around 0.5.   The 
system could be sealed and evacuated down to pressures as low as 0.7 
kPa.  The pump speed was limited to prevent air flow from disturbing the loose packing.   We checked for electrostatic charging by performing experiments in air at a high humidity ($\sim$50\%) where electrostatic effects typically vanish \cite{SLG06} and observed no qualitative change in the impact dynamics.

X-ray imaging was performed at the 
GSECARS  beamline at the Advanced Photon Source 
using a high intensity beam with energy 
width 5 keV  centered at 22 keV.   X-ray transmission through the bed was 
imaged off a phosphor screen at 6000 frames per second using a Phantom v7 video camera.  The beam size restricted the field of 
view to 22 mm x 8.7 mm sections of the bed. To capture the dynamics across 
the full vertical extent of the bed, movies of 
multiple independent drops, imaged at different, slightly overlapping 
vertical bed positions, were stitched together using the impacting sphere to align them horizontally and synchronize them.

The measured intensity, $I$, is a function of  the product 
$\rho \phi l$. Here $\rho$ is the density of the grain 
material and $l$ the x-ray path length through the bed, determined 
from the cylindrical geometry of the set-up.   To correct for spatial 
variations in beam intensity and  camera sensitivity, 
calibration curves relating  $I$ to the packing 
fraction $\phi$ were calculated for each of the 780 x 300 pixels in 
the field of view \cite{Rea07}.
Across each frame in a single movie, 
initial packing 
fraction $\phi_0$ varied by about 1\%, indicating a uniform bed packing prior to impact.   From drop to drop $\phi_0$ varied between 0.49 and 0.52.  This 
variation was present at atmospheric pressure, where the 
pump was disconnected, as well as at reduced pressure, indicating 
that it was due to small, unavoidable differences in bed settling 
after fluidization, but not due to the evacuation of the chamber.  

X-ray images of the interior
reveal a striking air pressure dependence of both bed and sphere dynamics (Fig. \ref{fig:xray}).  At 
atmospheric pressure ($P =$101 kPa), the sphere easily penetrates the 
bed, reaching the bottom of the system and opening up a large cylindrical hole. This hole closes first at some intermediate depth, resulting in a crater near the top and a gas-filled cavity behind the sphere. Both of these region then fill in from the sides due to gravitational pressure.  This process has previously been identified as driving jet formation \cite{Lea04, Rea05}, but the role of the interstitial gas has remained controversial \cite{CBMPL07}.
 At  $P =$12 kPa the overall features are still similar to those at atmospheric conditions, but the 
dynamics begin to change. The cavity is 
smaller, and while the top surface of the bed still rises to compensate for the change in bed 
volume, it does not rise quite as high (Fig. \ref{fig:xray}a-c).   Further reducing the 
pressure to 0.7 kPa significantly changes the dynamics.  The bed barely rises, and there is significant 
compaction in front of the sphere, evident in 
the darker region under the sphere in Fig. \ref{fig:xray}d-f. 
As a consequence, the sphere is able to penetrate only a short distance and the hole closes directly above the sphere without forming a separate cavity.

\begin{figure}
\includegraphics[width=7cm]{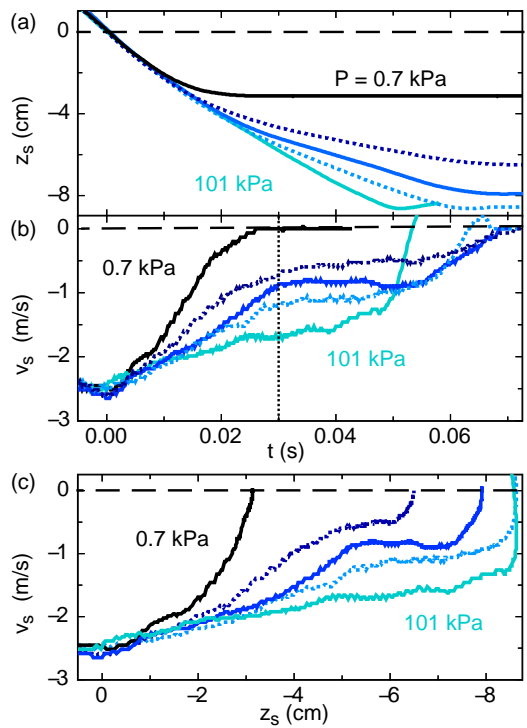}
\caption{(Color Online). (a) Sphere depth $z_s$ versus time after impact ($t = 0$ s). (b)  Velocity $v_s(t)$ computed from curves in (a).   (c)  Velocity $v_s (z_s)$ versus depth.   All panels top to bottom: $P = $ 0.7 kPa, 4.9 kPa, 8.7 kPa, 12 kPa and 101 kPa. }
\label{fig:ball_track}
\end{figure}   

To 
quantify this change in impact dynamics we plot in Fig. 
\ref{fig:ball_track}a the position of the bottom of the sphere, $z_s 
(t)$, as it moves through the bed.   At $P =$101 kPa the sphere 
hits the bottom with sufficient momentum to bounce back up a bit; at 
$P =$8.7 kPa and below the sphere is stopped well before reaching the 
bottom.  From $z_s (t)$ we  compute $v_s  = d 
z_s/dt$ (Fig. \ref{fig:ball_track} b, c).    With deceasing ambient pressure there is a monotonic decrease in penetration depth and an increase in bed resistance. 

Inside the bed the net force on the impacting projectile is the sum of its weight, $-mg$, and a drag force, $F_d$, representing the bed resistance. Models for this resistance are of the form $F_d = F_C + c|v_s|^\beta$, where $F_C$ represents Coulomb friction  and $c$ characterizes the strength of the velocity-dependent drag. Specific forms include $F_C = const$ with $\beta=1$ \cite{BW04},  $F_C = \kappa |z_s|$ with  $c = 0$ \cite{LRBM04, HPLLC05}, $F_C = \kappa |z_s|$ with $\beta=2$ \cite{KD07} as well as  $\beta=2$ but a more complex $z_s$-dependence for $F_C$ \cite{TV05}.   Note that all of these models predict a non-zero deceleration, $a=-g + F_d/m$, of the projectile and thus a $z_s$-dependence of its velocity over the whole range from impact to final stop.    

The data in Fig. \ref{fig:ball_track} for  intermediate pressures reveals a feature not captured in these models: a region of near-constant velocity beginning roughly 30 ms 
after the impact (dotted vertical line). This behavior can be seen most clearly when the velocity is plotted as function of depth, $z_s$, below the free surface (Fig. \ref{fig:ball_track}c). A second important feature is the rapid deceleration \emph{after} the constant velocity regime, seen clearly in the traces for $P=4.9$kPa and $8.7$kPa.  It is nearly as abrupt as when the sphere hits the bottom of the container (see traces for higher pressures), but occurs here sufficiently far inside the bed for boundary effects to be irrelevant \cite{Sea04}. As we show below, these characteristic features are closely linked to the interplay between penetrating sphere, bed particles, and interstitial gas.

\begin{figure}
\includegraphics[width=8cm]{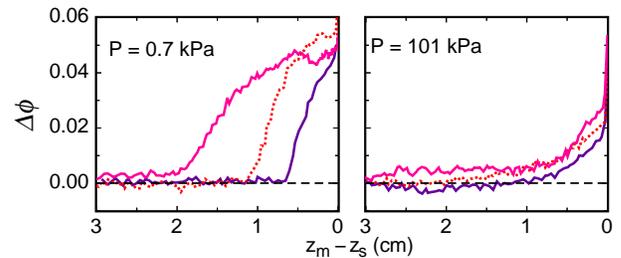}
\caption{(Color Online).  Compaction front 
preceding sphere.   Change in packing fraction $\Delta 
\phi$ measured along the path of the sphere at depths 
(left to right) $z_{m}$ = 1.0 cm, 2.0 cm and 3.0 cm 
plotted against distance from sphere tip $z_s$ to $z_{m}$.}
\label{fig:comp_front}
\end{figure}

\begin{figure}
\includegraphics[width=7.5cm]{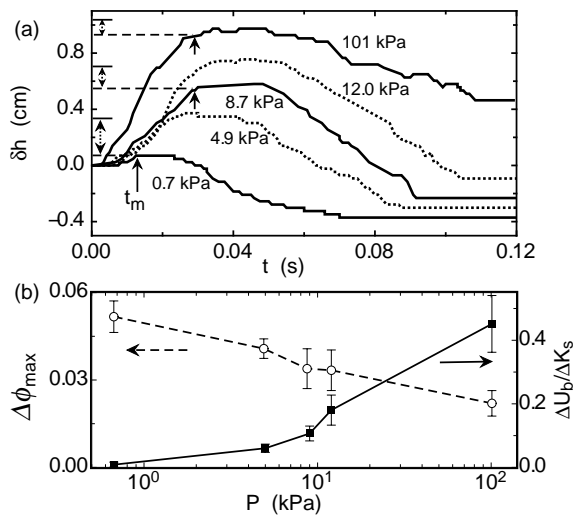}
\caption{Bed dynamics.  (a) Rise of bed surface $\delta h$.  Arrows mark $t_m$ for three pressures.  Double-sided arrows denote rise needed to conserve bed volume .  Resolution of $\delta h$ was limited by pixel size to  $\sim$0.5mm.  (b) Maximum 
change in packing fraction in front of sphere ($\bigcirc$) and ratio of potential energy needed to raise bed by $ \delta h_{max}$ 
to kinetic energy lost by sphere at time $t_m$ ($\blacksquare$).   Error bars for $\Delta \phi_{\textrm{max}}$  are due to fluctuations in $\Delta \phi$, and for $\Delta U_{b}/\Delta K_s$ due to uncertainty in $t_m$.
}
\label{fig:surf_rise}
\end{figure}

X-ray radiography allows us to examine this interplay locally.  In Fig. \ref{fig:comp_front}, we plot the change in 
local packing fraction, $\Delta \phi = \phi - \phi_0$, measured along 
the centerline of the path of the sphere, at three different depths, $z_m$ 
below the surface.  At $P=$ 0.7 kPa there is a clear jump in $\Delta \phi$  well before the 
sphere arrives. With 
increasing depth this jump occurs further ahead of the sphere, 
demonstrating that the compacted region grows as the sphere plows into grains that do not flow out of the way.   The situation is quite different at $P=$ 101 kPa.  The packing 
fraction remains constant except for a slow upturn once the sphere 
comes within about half its diameter of $z_m$. The width of this 
small compaction front varies little with depth. The overall magnitude of compaction decreases smoothly with increasing pressure (Fig. \ref{fig:surf_rise}c).       

A global measure of the effect of interstitial gas is the rise of the top surface as the sphere burrows into the bed. This rise can be seen  in Figs. \ref{fig:xray}a-c  but is considerably reduced when the system is evacuated (Figs. \ref{fig:xray}d-f).  In Fig. \ref{fig:surf_rise}a we track the level change, $\delta h$, of the top surface 
for different pressures. At impact, the bed rapidly rises, then at time  $t_m$ (highlighted for 101 kPa, 8.7 kPa and 0.7 kPa by arrows in Fig. \ref{fig:surf_rise}a)
levels off into a broad maximum of height $\delta h_{max}$ and eventually falls to a final level $\delta h < 0$  (at 101 kPa this final settling does not occur until times much later than shown).  The change in height  $\delta h_{max}$ increases with pressure, suggesting that the interstitial gas makes the bed as a whole less compressible and more fluid-like.

To check this, we use x-ray images as in Fig. \ref{fig:xray} to track the shape of the cavity and estimate its volume, assuming cylindrical symmetry. For atmospheric pressure we find that $\delta h(t)$ corresponds, within experimental uncertainties, to what would be expected from an incompressible fluid. With decreasing $P$ the level  $\delta h_{max}$ drops below the value obtained from the cavity volume (double-sided arrows in  Fig. \ref{fig:surf_rise}a), indicating a less elastic response.  

We can characterize the elasticity of the 
bed by comparing the gravitational energy gained by the bed $\Delta U_b$ to the 
kinetic energy lost by the sphere $\Delta K_s$ in time $t_m$.  While at $P = 
0.7$ kPa only 0.1\% of the impacting sphere's energy is transferred to the bed, this value increases to about 
45\% at atmospheric pressure  (Fig. \ref{fig:surf_rise}b). This can be compared to values around 10\% found in 2D simulations without interstitial gas \cite{TV05}. 
  
To understand the fluid-like behavior at large $P$,  we examine the rate of gas flow through the bed.  If the timescale for the expulsion of the gas from the bed is significantly longer than the timescale for the granular flow, then gas trapped and compressed by the bed can create pressure differences capable of supporting the bed \cite{Gu75, PVB95, MCKNJ04, Rea05}.  From Darcy's law and the continuity equation for the gas flow, one can derive a diffusion equation for the gas pressure $\frac{\partial P}{\partial t} = D \frac{\partial^2 P}{\partial^2 z}$, with diffusion constant $D = \frac{kP}{\mu (1-\phi)}$, where $\mu$ is gas viscosity and $k$ bed permeability \cite{Gu75, PVB95}.  For our experimental conditions  $D \sim 5$ cm$^2$/s at $P=101$ kPa.   The timescale to diffuse across the depth of our bed (a distance $L =$ 8.5 cm) is $\tau_D = L^2/D \sim$ 
140 ms, significantly longer than the time $t_m \sim$ 30 ms for the bed to rise to $\delta h_{max}$.  This suggests that air trapped in the bed interstices prevents compaction at large $P$.   Since the permeability depends on the grain diameter 
according to $k\sim d^2$, this cushioning effect would be less pronounced with 
larger grains, as noted in \cite{GTBZM99}.  Conversely, we expect the behavior of larger grains  to resemble that found at our lowest pressures.  Indeed, the  trajectory of the sphere at 0.7 kPa is qualitatively similar to trajectories measured by Durian et al. in 250$\mu$m - 350$\mu$m glass spheres \cite{UAOD03, KD07}.  

The air-mediated response of the granular bed directly affects the drag on the impacting sphere. Comparing Figs. \ref{fig:surf_rise}a and  \ref{fig:ball_track}b,c we see that the onset of the constant velocity regime coincides with $t_m$, the start of the plateau in bed level rise.  During this stage the material displaced by the sphere is flowing mostly into the cavity behind 
it.  Consequently, the sphere is not affected by the full bed, but instead by a more local region. For $v_s \sim 1$ m/s, as for our data, the 
drag $\phi \rho D_s^2 v_s^2$ is within a factor of two of the weight of the sphere.  This suggests a large reduction of  $F_C$ in the force law in the observed constant velocity regime. 
When the top level of the bed begins to 
fall again,  the cavity behind the sphere has pinched shut (Fig. 
\ref{fig:xray}c), trapping an air pocket below the surface 
\cite{Rea05}.  With the falling bed and trapped air pocket, the bed material is no longer free to flow out of 
the way of the sphere.   Coulomb friction is again set by 
the full weight of the bed, resulting in an increase in $F_C$ that quickly decelerates the sphere and brings it to rest.   

As a result, a friction term of the form  $F_C= \kappa|z|$ cannot capture the full range of observed behavior, even if $\kappa$ is made to depend on pressure.  Such pressure dependence was very recently proposed by Cabarello \textit{et al.} 
\cite{CBMPL07} who measured the trajectory of a sphere impacting 
40 $\mu$m sand by tracking a string attached to it.  Our results agree with their conclusion that the shallower penetration at 
lower pressures is due to increased friction. However, since $\Delta \phi>0$ (Fig. \ref{fig:comp_front}) at all pressures, drag reduction is not simply due to  fluidization of grains ahead of the sphere.  Instead, our 
measurements of the local packing fraction indicate that in 
the presence of air the bed as a whole behaves more like an incompressible fluid, 
allowing the impacting sphere to penetrate deep and  create a large cavity.   In the absence of air the bed compacts much more strongly ahead of 
the sphere, rapidly dissipating energy and decelerating the descent.   

We thank B. Conyers, M. M\"{o}bius, S. Nagel and M. River for insightful discussions, and 
D. Durian and D. Lohse for pre-publication results. 
This work was supported by NSF through MRSEC DMR-02137452 and by DoD/AFOSR FA9550-07-101459. GSECARS was supported by NSF EAR-0217473, DOE DE-FG02-94ER14466, the State of Illinois and the W. M. Keck Foundation, and use of the Advanced Photon Source by DOE-BES W-31-109-Eng-38.

\end{document}